%% file: main_v5.tex
\documentclass[a4paper, 11pt]{article}
\pdfoutput=1
\usepackage{jheppub}
\usepackage[colorlinks = true, citecolor = blue, linkcolor = blue, urlcolor = blue]{hyperref}

\input{packages}

\input{definitions}

\usepackage{graphicx}
\usepackage{tabularx}
\usepackage{physics}
\usepackage[normalem]{ulem}
\usetikzlibrary{shapes.symbols}
\usepackage{tikz}
\usepackage{mathtools}
\usepackage[capitalise]{cleveref} 
\usepackage{bm}
\usepackage{mathrsfs}
\usepackage{xcolor}
\usepackage{diagbox}
\usepackage{subfigure}
\definecolor{tangerine}{RGB}{242, 133, 0}
\definecolor{persianGreen}{RGB}{0, 166, 147}

\newcommand{\update}[1]{{#1}}

\title{Probing the Sivers Asymmetry with Transverse Energy-Energy Correlators in the Small-$x$ Regime}

\author[a]{Shohini Bhattacharya,}
\author[b,c,d]{Zhong-Bo Kang,}
\author[b,c]{Diego Padilla,}
\author[b,c]{and Jani Penttala}

\affiliation[a]{Department of Physics, University of Connecticut, Storrs, CT 06269, USA}
\affiliation[b]{Department of Physics and Astronomy, University of California, Los Angeles, CA 90095, USA}
\affiliation[c]{Mani L. Bhaumik Institute for Theoretical Physics, University of California, Los Angeles, CA 90095, USA}
\affiliation[d]{Center for Frontiers in Nuclear Science, Stony Brook University, Stony Brook, NY 11794, USA}

\emailAdd{shohinib@uconn.edu}
\emailAdd{zkang@physics.ucla.edu}
\emailAdd{dpadi022@g.ucla.edu}
\emailAdd{janipenttala@physics.ucla.edu}

\abstract{We investigate transverse energy--energy correlators (TEECs) for both polarized and unpolarized targets in the small-$x$ regime at the Electron-Ion Collider (EIC). Focusing on the approximately back-to-back electroproduction of a hadron--electron pair, we apply transverse-momentum-dependent (TMD) factorization formulas that incorporate TMD evolution for both event-shape observables and expand them in terms of the small-$x$ dipole amplitude.
This allows us to write the TEEC off the transversely polarized proton in terms of a C-odd interaction, corresponding to an odderon exchange.
Due to the charge-conjugation-odd nature of the small-$x$ quark Sivers function, we restrict the sum over final hadronic states to positively and negatively charged hadrons separately.
We present numerical predictions for the TEEC Sivers asymmetry at the EIC and find the magnitude of the asymmetry to be on the $0.01 \%$ level.
This channel offers a promising avenue for benchmarking the still largely unconstrained odderon amplitude.}

\begin{document}

\maketitle

\section{Introduction} \label{sec:intro}

Understanding the transverse momentum structure of partons and the behavior of nuclear matter at small values of Bjorken $x$ are central goals in modern QCD. Together, these areas offer complementary insights into the three-dimensional imaging of the proton and the emergent dynamics of high-density gluon fields, including the onset of gluon saturation~\cite{Boer:2011fh,Iancu:2003xm}. Achieving simultaneous access to both the transverse and small-$x$ regimes requires a facility with high luminosity, polarization control, and broad kinematic reach. The forthcoming Electron-Ion Collider (EIC) is designed to meet these demands~\cite{AbdulKhalek:2021gbh,AbdulKhalek:2022hcn,Accardi:2012qut}, offering a unique opportunity for precision studies of partonic structure in nucleons and nuclei across a wide range of energies and momentum fractions.

There has been a recent interest in studying the overlap between these two areas in the context of the study of transverse-momentum-dependent (TMD) parton distribution functions~\cite{Boussarie:2023izj} in the dipole picture of high-energy scattering~\cite{Zhu:2024iwa,Benic:2024fbf,Kovchegov:2022kyy,Kovchegov:2021iyc,Boussarie:2019vmk,Dong:2018wsp,Hatta:2022bxn,Chen:2024bpj,Caucal:2025qjg,Caucal:2025xxh}.
Additionally, since the measurement of the in-jet energy--energy correlators (EEC)~\cite{Komiske:2022enw,Lee:2022uwt}, there has been a comeback of these event-shape observables for high-energy QCD studies~\cite{Kang:2023oqj,Kang:2024otf,Kang:2024dja,Barata:2024wsu,Mantysaari:2025mht,CMS:2025ydi,STAR:2025jut,Craft:2022kdo,Devereaux:2023vjz,Kang:2023big,Lee:2023npz,Lee:2023tkr,Lee:2024esz,Kang:2025vjk}.
In this work, we follow the spirit of Ref.~\cite{Kang:2023oqj}, where an event-shape observable is formulated within the TMD factorization framework and subsequently expanded in terms of the dipole amplitude from the color glass condensate (CGC) effective theory~\cite{Gelis:2010nm,Weigert:2005us}, which describes high-energy QCD dynamics at small Bjorken-$x$.

We are particularly interested in the quark Sivers function, which was first introduced in Refs.~\cite{Sivers:1989cc,Sivers:1990fh} to explain the large single-spin asymmetries of pion production in hadron--hadron scattering.
This TMD encodes the quantum correlation between the proton spin and the intrinsic motion of quarks, and
it can be interpreted as the number density of unpolarized partons inside a transversely polarized proton.
\update{
While the quark Sivers function has been studied extensively before~\cite{Echevarria:2020hpy,Cammarota:2020qcw,Fernando:2023obn,Bacchetta:2020gko,Bury:2020vhj,Liu:2018trl}, it nevertheless remains not well constrained for the small-$x$ region.}

This work focuses on providing a new channel to probe the Sivers asymmetry at small $x$ through the transverse energy--energy correlator (TEEC)~\cite{Ali:1984yp} which can be measured at the EIC.
The TEEC observable is a generalization of the EEC that is more suitable for hadronic colliders.
\update{They have the advantage, like other energy correlators, of being widely inclusive by involving a sum over hadronic final states. This reduces the dependence on the non-perturbative hadronization of the final partons compared to the conventional azimuthal correlations associated with identified hadrons~\cite{Ganguli:2025aqa}, while the energy weighting suppresses contributions from very soft hadrons. Moreover, compared with jet observables, such as the lepton--jet correlations~\cite{Liu:2018trl,Arratia:2020nxw}, TEECs do not require explicit jet reconstruction and therefore avoid dependence on the jet definition and associated jet dynamics~\cite{Hatta:2021jcd,Kang:2026zmf}. These features make them well-suited to probing initial-state distributions.}
For this work, however, we will focus on working with TEECs that are disjointedly inclusive in positively and negatively charged hadrons, which will be key in obtaining a non-vanishing Sivers asymmetry.
This corresponds to defining the TEEC with charge tracks~\cite{Lee:2023npz,Chen:2020vvp,Lee:2023tkr,Li:2021zcf,Jaarsma:2023ell,Kang:2023big,Mantysaari:2025mht}.

\update{
The small-$x$ behavior of the Sivers function is particularly interesting in the context of CGC due to its dependence on the imaginary part of quark dipole $S$-matrix which corresponds to the spin-dependent odderon amplitude \cite{Hatta:2005as,Yao:2018vcg,Dong:2018wsp,Bhattacharya:2023yvo,Kovchegov:2021iyc,Zhu:2024iwa,Mantysaari:2025mht,Zhou:2013gsa}.
}
For comparison, the unpolarized quark TMD depends on the real part of the quark dipole $S$-matrix, the so-called pomeron term, which involves a C-even interaction with the target proton.
In the small-$x$ limit, the pomeron term dominates, and such interaction is well understood using the McLerran--Venugopalan model~\cite{McLerran:1993ka,McLerran:1994vd,McLerran:1993ni}.
In contrast, the odderon amplitude remains poorly understood
due to its subleading nature,
and the existence of the odderon interaction has been demonstrated experimentally only quite recently~\cite{D0:2020tig}.
In this paper, we will discuss how the Sivers asymmetry is directly related to the odderon amplitude, and therefore our proposed observable serves as a testing ground for benchmarking the odderon amplitude at small $x$.

The paper is structured as follows.
In Sec.~\ref{sec:Theory formalism}, we give an overview of the TEEC factorization for both the unpolarized and polarized cases and go over the small-$x$ expansion and evolution.
In Sec.~\ref{sec:Pheno}, we present our numerical predictions using the formalism developed, focusing on the Sivers asymmetry in the TEEC observable at the future EIC. Finally, we summarize our findings and outlook in Sec.~\ref{sec:Conclusion}.

\section{Theoretical Formalism} \label{sec:Theory formalism}
\begin{figure}[t]
    \centering
    \includegraphics[scale=1.3]{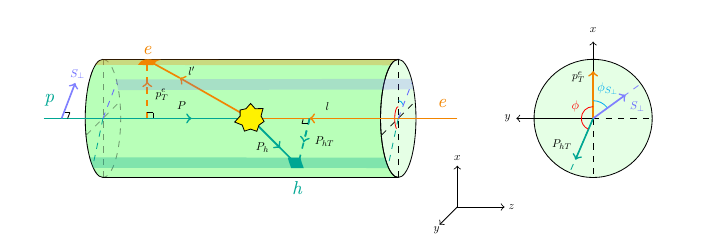}
    \caption{Left panel: Illustration of the TEEC for DIS off a transversely polarized proton in the lab frame, where the incoming proton defines the positive $z$-axis.
    Right panel: Transverse plane, where the x-axis is defined by the transverse momentum of the scattered electron. }
    \label{fig:Diagram}
\end{figure}

We study the TEEC between the outgoing electron and the produced hadrons in deep inelastic scattering (DIS), following the framework of Refs.~\cite{Kang:2023oqj,Li:2020bub}. The process of interest is
\begin{align}
    e(l) + p(P,\,S_\perp) \rightarrow e(l') + h(P_h) + X\,,
\end{align}
where $l$ and $l'$ are the four-momenta of the incoming and outgoing electron, with $q^2 = (l' - l)^2 = -Q^2$ denoting the virtuality of the exchanged photon. The four-momentum and transverse spin of the incoming proton are denoted by $P$ and $S_\perp$, respectively, and $P_h$ is the momentum of the final-state hadron. We work in the center-of-mass frame of the electron--proton system, where the proton (electron) moves along the $+z$ ($-z$) direction, and consider the back-to-back limit, where the produced hadron and the scattered electron are nearly opposite in the transverse plane, as illustrated in Fig.~\ref{fig:Diagram}.

The TEEC observable in unpolarized electron--proton scattering was studied in Ref.~\cite{Kang:2023oqj}, where the effects of gluon saturation in the small-$x$ regime were incorporated. The TEEC is a transverse-energy-weighted cross section measured as a function of the azimuthal angle $\phi$ between the outgoing electron and the hadron. It is defined as~\cite{Li:2020bub}
\begin{align}
\frac{\dd \Sigma}{\dd\xi}
\equiv \sum_{h}\int \dd E_{T,l'}\, \dd E_{T,h}\, \dd\phi\,
\frac{\dd{\sigma}}{\dd\phi\, \dd E_{T,l'} \,\dd E_{T,h}}
  \frac{ E_{T,l'} E_{T,h}}{E_{T,l'} \sum_{h'}E_{T,h'}}\delta\left(\xi- \left( \pi-\phi \right)\right) \,,
  \label{eq:TEEC}
\end{align}
where $E_{T,i} = \sqrt{p_{T,i}^2 + m_i^2} \approx p_{T,i}$ is the transverse energy of particle $i$ (see Fig.~\ref{fig:Diagram} for an illustration of the coordinate system). The variable $\xi$\footnote{{We choose to work with $\xi$ rather than $\tau = \frac{1+\cos{\phi}}{2}$ \cite{Kang:2023oqj, Li:2020bub}. This choice is more convenient because, as we will show below, the Sivers asymmetry in the TEEC is an odd function of $\xi$ for a polarized proton.}} is defined as
\begin{align}
    \xi = \pi -\phi \,,
\end{align}
such that the back-to-back limit $\phi \approx \pi$ corresponds to $|\xi | \ll 1$.

In this study, we extend the TEEC observable to the case of polarized electron--proton scattering. In particular, we consider the scenario in which the incoming proton beam is transversely polarized, and investigate the associated transverse spin asymmetry---commonly referred to as the Sivers asymmetry---in the TEEC observable. When the proton is transversely polarized, a correlation arises between its transverse spin vector $S_\perp$ and the intrinsic transverse momentum of the quark inside the proton, as described by the quark Sivers function. This correlation can induce a preferred transverse orientation of the final-state hadron, leading to a characteristic azimuthal modulation in the TEEC observable of the form $\propto \cos(\phi_{S_\perp} - \phi_{l'})$, where $\phi_S$ and $\phi_{l'}$ denote the azimuthal angles of the proton spin and the outgoing electron, respectively. We can express the spin-dependent TEEC observable in the polarized electron--proton scattering as
\begin{equation}
    \frac{\dd \Sigma}{\dd\xi\,\dd y_{e}\,\dd^{2}\mathbf{p}^{e}_{T}}
    = \Sigma^{UU} + \cos(\phi_{S_\perp} - \phi_{l'})\,\Sigma^{UT} \,,
    \label{eq:Sigma-full}
\end{equation}
where $\Sigma^{UU}$ denotes the unpolarized contribution, and $\Sigma^{UT}$ encodes the transverse spin--dependent contribution associated with the Sivers effect. Here, $y_e$ and $p^{e}_{T}$ are the rapidity and transverse momentum of the final-state electron.

The standard TEEC observable involves a sum over all final-state hadrons, as indicated by the $\sum_h$ in Eq.~\eqref{eq:TEEC}. In this study, we generalize the TEEC to consider only a subset $\mathbb{S}$ of hadrons---for example, restricting the sum to either positively or negatively charged hadrons. Less inclusive versions of EEC observables have also been explored in Ref.~\cite{Kang:2023big}, where the so-called Collins-type EEC was introduced to study spin effects in the final state. Experimentally, such selections are feasible: for instance, the ALICE Collaboration at the LHC has measured the EEC using only charged particles~\cite{ALICE:2024dfl}. This selective treatment becomes particularly relevant in the context of the Sivers asymmetry in the TEEC at small $x$, as we will discuss below. To reflect this generalization, we introduce a subscript $\mathbb{S}$ on the TEEC observable and denote it as $\Sigma \to \Sigma_{\mathbb{S}}$ throughout the remainder of this paper.

\update{
In the back-to-back limit, the  TEEC can be factorized within the TMD framework in terms of a small transverse acoplanarity, see e.g. Refs.~\cite{Li:2020bub,Kang:2023oqj,Gao:2023ivm}. Here we use the same factorization but keep our observable differential in the signed angular variable $\xi$ as discussed above. This can be written as
\begin{align}
    \frac{d\Sigma_{\mathbb{S}}}{d\xi\,dy^e\,d^2p_T^e} &= \sigma_0H\left(Q,\mu\right)\sum_{i=q,\bar{q}}e_I^2p^e_T\sum_{h\in\mathbb{S}}\int_0^1
dz\,z\int dp_y\,dk_y\,d\lambda_y\,
\delta\left(p_T^e\xi -p_y-k_y/z+\lambda_y\right) \nonumber \\
&\quad\times  S_{nn_h}(\lambda_y,\mu,\nu)\,\hat{f}_{i/p}^{\left(u\right)}(x,p_y,\mu,\zeta/\nu^2)\,D^{\left(u\right)}_{1,h/i}(z,k_y,\mu,\zeta'/\nu^2). \label{eq:startFact}
\end{align}
where $H(Q,\mu)$ is the hard function for DIS, $D^{\left(u\right)}_{h/q}(z,k_y,\mu,\zeta'/\nu^2)$ is the unsubtracted and unpolarized quark TMD FF, $\hat{f}_{q/P}^{\left(u\right)}(x,p_y,\mu,\zeta/\nu^2)$ is the unsubtracted quark TMD PDF and $S_{nn_h}(\lambda_y,\mu,\nu)$ is the soft function relevant for this process~\cite{Li:2020bub}. The scales $\mu$, $\zeta$, and $\zeta'$ are the renormalization and Collins--Soper (CS) scales that govern the evolution of the TMDs and $\nu$ is the rapidity scale which cancels between the soft function and the TMD PDF and TMD FF.

Note that the momentum imbalance in Eq.~\eqref{eq:startFact} is one dimensional and therefore the distributions we have are given by \cite{Gao:2023ivm} $ F(l_y) = \int dl_x \,F(l_\perp)$ where in the right hand side we have the usual two dimensional distribution. The quark TMD PDF can be separated into unpolarized and polarized pieces as~\cite{Boussarie:2023izj}
\begin{align}
    \hat{f}_{q/p}^{\left(u\right)}(x,p_y,\mu,\zeta/\nu^2)&=  \int dp_x \left[f^{\left(u\right)}_{q/p}(x,p_\perp,\mu,\zeta/\nu^2)+\frac{\epsilon_T^{\rho \sigma}p_{\perp\rho}S_{T\sigma}}{M}f_{1T,q/p}^{\left(u\right)\perp}(x,p_\perp,\mu,\zeta/\nu^2) \right] \nonumber \\ &=
f_{q/p}^{\left(u\right)}\left(x,p_y,\mu,\zeta/\nu^2\right) -\frac{p_y\cos(\phi_{S_\perp} - \phi_{l'})}{M}f_{1T,q/p}^{\left(u\right)\perp}\left(x,p_y,\mu,\zeta/\nu^2\right), \label{eq:FT_TMDPDF}
\end{align}
where the first term is the unpolarized quark TMD PDF and the second spin-dependent term involves the Sivers function. $S_{\perp}$ is the transverse spin vector of the proton, $M$ is the proton mass, and $\epsilon_T^{\mu\nu}$ is the two-dimensional Levi--Civita symbol.

To simplify Eq.~\eqref{eq:startFact} we use the integral representation of the delta function and the position space form of the distributions~\footnote{We note that the TMD fragmentation function $D_{1,h/q}\left(z,b,\mu,\zeta'\right)$ used here follows a slightly different convention from that in the TMD Handbook~\cite{Boussarie:2023izj}, in particular omitting the usual $1/z^2$ prefactor.}
\begin{equation}
\tilde{f}^{\left(u\right)}_{q/p}(x,b_y,\mu,\zeta/\nu^2) = \int dp_y \, e^{-ip_yb_y} \,f^{\left(u\right)}_{q/p}(x,p_y,\mu,\zeta/\nu^2), \label{eq:FT_unpol}
\end{equation}
\begin{equation}
i M^2b_y \tilde{f}_{1T,q/p}^{\left(u\right)\perp}\left(x,b_y,\mu,\zeta/\nu^2\right) = \int dp_y \, e^{-ip_yb_y} \,p_y f_{1T,q/p}^{\left(u\right)\perp}\left(x,p_y,\mu,\zeta/\nu^2\right), \label{eq:FTSivers}
\end{equation}
\begin{equation}
\tilde{S}_{nn_h}(b_y,\mu,\nu) = \int d\lambda_y \, e^{i\lambda_yb_y} S_{nn_h}(\lambda_y,\mu,\nu) ,
\end{equation}
\begin{equation}
    \tilde{D}^{\left(u\right)}_{1,h/q}(z,b_y,\mu,\zeta'/\nu^2) = \int  d k{_y}\, e^{-i k_yb_y/z}\, D^{\left(u\right)}_{1,h/q}(z,k_y,\mu,\zeta'/\nu^2)\label{eq:FT_FF} ,
\end{equation}
where the convention in Eq.~\eqref{eq:FTSivers} is chosen such that $\tilde{f}_{1T}^{\perp}(x,b_y)$ is unit-less and the convention in Eq.~\eqref{eq:FT_FF} follows the work in \cite{Kang:2023oqj}. Together this gives
\begin{align}
\Sigma^{UU}_{\mathbb S}
={}& \sigma_0 H(Q,\mu)\sum_{i=q,\bar q} e_I^2 p_T^e
\int_{-\infty}^{\infty}\frac{db_y}{2\pi}\,
e^{i\xi p_T^e b_y}\,
\tilde S_{nn_h}(b_y,\mu,\nu)
\notag\\
&\times
\tilde f_{i/p}^{(u)}(x,b_y,\mu,\zeta/\nu^2)
\sum_{h\in\mathbb S}\int dz\,z\,
\tilde D_{1,h/i}^{(u)}(z,b_y,\mu,\zeta'/\nu^2)
\notag\\
={}& \sigma_0 H(Q,\mu)\sum_{i=q,\bar q} e_I^2 p_T^e
\int_{-\infty}^{\infty}\frac{db_y}{2\pi}\,
e^{i\xi p_T^e b_y}\,
\tilde f_{i/p}(x,b_y,\mu,\zeta)
\notag\\
&\times
\sum_{h\in\mathbb S}\int dz\,z\,
\tilde D_{1,h/i}(z,b_y,\mu,\zeta') ,
\label{eq:partialSimgUU}
\end{align}
\begin{align}
\Sigma^{UT}_{\mathbb{S}}
={}& -iM\sigma_0 H(Q,\mu)\sum_{i=q,\bar q} e_I^2 p_T^e
\int_{-\infty}^{\infty}\frac{db_y}{2\pi}\,
e^{i\xi p_T^e b_y}\,b_y\,
\tilde S_{nn_h}(b_y,\mu,\nu)
\notag\\
&\times
\tilde f_{1T,i/p}^{(u)\perp}(x,b_y,\mu,\zeta/\nu^2)
\sum_{h\in\mathbb S}\int dz\,z\,
\tilde D_{1,h/i}^{(u)}(z,b_y,\mu,\zeta'/\nu^2)
\notag\\
={}& -iM\sigma_0 H(Q,\mu)\sum_{i=q,\bar q} e_I^2 p_T^e
\int_{-\infty}^{\infty}\frac{db_y}{2\pi}\,
e^{i\xi p_T^e b_y}\,b_y\,
\tilde f_{1T,i/p}^{\perp}(x,b_y,\mu,\zeta)
\notag\\
&\times
\sum_{h\in\mathbb S}\int dz\,z\,
\tilde D_{1,h/i}(z,b_y,\mu,\zeta') ,
\label{eq:partialSimgUT}
\end{align}
where we have absorbed the soft function into the unsubtracted TMD PDF and TMD FF, which is done in detail in \cite{Kang:2023oqj}. We can now identify the TEEC jet function $J_{\mathbb{S}/q}$ via
\begin{align}
J_{\mathbb{S}/q}\left(b_y,\mu,\zeta'\right) \equiv \sum_{h\in \mathbb{S}}\int_{0}^{1}\dd{z}   z \tilde{D}_{1,h/q}\left(z,b_y,\mu,\zeta'\right)\,,
\label{eq:jetfun1}
\end{align}
after which we obtain
\begin{equation}
    \Sigma^{UU}_{\mathbb{S}}=\sigma_{0} H(Q,\mu) \sum_{i = q, \bar q} e_{i}^{2} p_{T}^{e} \int_{0}^{\infty} \frac{\dd b}{\pi} \cos\left(\xi b p_{T}^{e} \right) f_{i/p}\left(x,b,\mu,\zeta\right) J_{\mathbb{S}/i}\left(b,\mu,\zeta'\right) \,, \label{eq:TEECsub}
\end{equation}

and
\begin{equation}
    \Sigma_{\mathbb{S}}^{UT}  = \sigma_{0} H(Q,\mu)\,
    M \sum_{i = q, \bar q} e_{i}^{2} p_{T}^{e}
    \int_{0}^{\infty} \frac{\dd b}{\pi}
    \sin\left(\xi bp_{T}^{e}\right)\,
    b\, f_{1T,i/p}^{\perp}(x,b,\mu,\zeta)\, J_{\mathbb{S}/i}(b,\mu,\zeta').\label{eq:TEEC_polarized}
\end{equation}
For simplicity of notation, we have dropped the tilde in the position-space distributions, as well as the $y$ subscript in the convolution coordinate $b_y$, since the latter is now restricted to positive values in the integral. Note that the appearance of the additional factor ``$i\,b$'' in the substitution leads to a $\sin\left(\xi bp_{T}^{e}\right)$ dependence in the Sivers case, in contrast to the $\cos\left(\xi bp_{T}^{e}\right)$ term in Eq.~\eqref{eq:TEECsub} for the unpolarized contribution. The resulting azimuthal angle dependence on the spin, $\cos(\phi_{S_\perp} - \phi_{l'})$, has been explicitly singled out in Eq.~\eqref{eq:Sigma-full}. This angular modulation is consistent with previous studies of the Sivers asymmetry; see, e.g., Refs.~\cite{Kang:2023big, Gao:2022bzi}.
}

Throughout this work, we choose $\mu^2 = \zeta = \zeta' = Q^2$, which satisfies the renormalization group consistency condition $\zeta \zeta' = Q^4$~\cite{Kang:2023oqj}. The overall normalization factor $\sigma_0$ is the leading-order partonic electron--quark cross section, given by
\begin{align}
    \sigma_{0} = \frac{2\alpha_{\text{em}}}{s Q^2} \frac{\hat{s}^2 + \hat{u}^2}{\hat{t}^2} \,,
\end{align}
where the partonic Mandelstam variables are defined as
\begin{align}
    \hat{s} &= x s \,, \\
    \hat{t} &= -Q^2 = -p_T^e e^{y_e} \sqrt{s} \,, \\
    \hat{u} &= -\hat{s} - \hat{t} \,.
\end{align}
Here $s$ is the center-of-mass energy of the electron--proton system. The momentum fraction $x$ is related to the outgoing electron’s kinematics in the back-to-back limit by
\begin{align}
    x = \frac{p_{T}^{e} e^{y_{e}}}{\sqrt{s} - p^{e}_{T} e^{-y_{e}}} \,.
\end{align}
We define the Sivers asymmetry as the ratio of the polarized to the unpolarized contributions to the TEEC observable in electron--proton scattering:
\begin{align}
    A_{UT}^{\mathbb{S}} = \Sigma_{\mathbb{S}}^{UT} \left/\, \Sigma_{\mathbb{S}}^{UU} \,\right. .
    \label{eq:Asymmetry}
\end{align}

\subsection{Quark TMDs in the small-$x$ regime}\label{sec:small-x}
The quark TMDs obey evolution equations that describe their dependence on the renormalization scale $\mu$ and the Collins--Soper (CS) scale $\zeta$. In coordinate space, these evolution equations diagonalize and can be solved in the perturbative region, where $b \lesssim 1/\Lambda_{\text{QCD}}$. To describe the non-perturbative region, we employ the $b_*$-prescription, which enables a smooth interpolation between perturbative and non-perturbative physics~\cite{Kang:2023oqj,Kang:2024otf,Echevarria:2020hpy,Sun:2014dqm,Isaacson:2023iui,Collins:1984kg}. This leads to the following expressions for the evolved TMDs:
\begin{align}
    f_{q/p}(x,b,\mu,\zeta) &= f_{q/p}(x,b,\mu_{b_{*}},\mu_{b_{*}}^{2})\, \exp\left[-S_{\text{NP}}(b,Q_{0},\zeta)\right] \exp\left[-S_{\text{pert}}(\mu,\mu_{b_{*}},\zeta)\right] \,, \label{eq:EvolvedTMDunpol1} \\
    f_{1T,q/p}^{\perp}(x,b,\mu,\zeta) &= f_{1T,q/p}^{\perp}(x,b,\mu_{b_{*}},\mu_{b_{*}}^{2})\, \exp\left[-S^{s}_{\text{NP}}(b,Q_{0},\zeta)\right] \exp\left[-S_{\text{pert}}(\mu,\mu_{b_{*}},\zeta)\right] \,,
    \label{eq:EvolvedSivers1}
\end{align}
where $\mu_{b_{*}} = 2e^{-\gamma_{E}}/b_{*}$ and $b_{*} = b / \sqrt{1 + b^{2} / b^{2}_{\max}}$. Following Refs.~\cite{Echevarria:2020hpy,Kang:2023oqj,Kang:2024otf}, we take $b_{\max} = \SI{1.5}{GeV^{-1}}$. The perturbative Sudakov factor $S_{\text{pert}}$ is given by
\begin{align}
S_{\text{pert}}\left(\mu, \mu_{b_*}, \zeta\right)
=&
- K \left(b_*, \mu_{b_*} \right) \ln\left( \frac{\sqrt{\zeta}}{\mu_{b_*}} \right)
- \int_{\mu_{b_{*}}}^{\mu} \frac{\dd{\mu'}}{\mu'}\, \gamma_{\mu}^q \left[\alpha_s(\mu'), \frac{\zeta}{\mu'^2} \right] \,,
\label{eq:perturbative_Sudakov}
\end{align}
where $K(b, \mu)$ is the CS evolution kernel~\cite{Collins:2011zzd,Boussarie:2023izj,Moult:2022xzt,Duhr:2022yyp,Vladimirov:2020umg}, and $\gamma_\mu^q$ is the anomalous dimension. Throughout this work, we employ next-to-leading logarithmic (NLL) accuracy; see, e.g., Refs.~\cite{Kang:2024otf,Echevarria:2020hpy}
for explicit expression for $K(b, \mu)$ and $\gamma_\mu^q$.

The non-perturbative Sudakov factors $S_{\text{NP}}$ and $S^s_{\text{NP}}$ account for the behavior of TMDs at large values of $b$ and must be modeled. For example, Ref.~\cite{Sun:2014dqm} uses the following parametrization for the unpolarized quark TMD:
\begin{equation}
    S_{\text{NP}}\left(b,Q_{0},\zeta\right) = \frac{g_{2}}{2} \ln\left(\frac{\sqrt{\zeta}}{Q_{0}}\right) \ln\left(\frac{b}{b_{*}}\right) + g_{1} b^{2} \,,
\end{equation}
with $g_{2} = 0.84$, $Q_{0}^{2} = \SI{2.4}{GeV^2}$, and $g_{1} = \SI{0.106}{GeV^2}$. Similarly, Ref.~\cite{Echevarria:2020hpy} uses the following parametrization in the extraction of the Sivers function:
\begin{equation}
     S^{s}_{\text{NP}}\left(b,Q_{0},\zeta\right) = \frac{g_{2}}{2} \ln\left(\frac{\sqrt{\zeta}}{Q_{0}}\right) \ln\left(\frac{b}{b_{*}}\right) + g_{1}^{s} b^{2} \,,
\end{equation}
where the only difference from the unpolarized case lies in the coefficient $g_{1}^{s} = \SI{0.180}{GeV^2}$. The term proportional to $g_2$ in both $S_{\text{NP}}$ and $S^s_{\text{NP}}$ originates from the non-perturbative component of the CS kernel and is spin-independent, reflecting universal $\zeta$-evolution. In contrast, the terms $g_1 b^2$ and $g_1^s b^2$ describe the intrinsic transverse motion of quarks inside the proton and differ between the unpolarized and Sivers cases.

In the standard TMD modeling, one typically expands the quark TMDs at the initial scale $\mu_0^2=\zeta_0 = \mu_{b_*}^2$ in terms of the collinear quark and gluon distribution functions $f_{q,g/p}(x, \mu_{b_*})$ via the operator product expansion. In this paper, however, we follow Ref.~\cite{Kang:2023oqj} and instead model the quark TMDs using the CGC effective field theory. Accordingly, we expand the quark TMDs at the initial scale in terms of dipole amplitudes in the CGC framework~\cite{Marquet:2009ca,Tong:2022zwp,Dong:2018wsp}: \update{
\begin{multline}
    \hat{f}_{q/p}(x,b,\mu_{b_{*}},\mu_{b_{*}}^{2}) \overset{\text{small } x}{=} \frac{N_{c}B_{\perp}}{8\pi^4}\frac{1}{x}\int \dd[2]{r}  \dd{\epsilon^{2}_{f}} \mathcal{K}\left(\bold{b},\bold{r},\epsilon_f
    \right)\big( 2-S_{x}\left(\mathbf{r+b}\right)-S_{x}\left(-\mathbf{r}\right)\big)\,,
    \label{eq:TMD_x_expanded}
\end{multline}
where $\mathcal{K}\left(\bold{b},\bold{r},\epsilon_f\right)$ is a perturbatively calculable kernel given by:
\begin{equation}
    \mathcal{K}\left(\bold{b},\bold{r},\epsilon_f\right) = \mathbf{\frac{\left(b+r\right)\cdot r}{|b+r||r|}}\epsilon^{2}_{f}  \, K_{1}\left(\epsilon_{f}\mathbf{|b+r|}\right) K_{1}\left(\epsilon_{f}\mathbf{|r|}\right).\label{eq:Kernel}
\end{equation}}
\update{
Here $\hat{f}_{q/p}\left(x,b,\mu_b^*,\mu_b^{*2}\right)$ denotes the (subtracted) position space quark distribution inside a transversely polarized proton and can be decomposed as
\begin{align}
            \hat{f}_{q/p}\left(x,b,\mu,\zeta\right) = f_{q/p}\left(x,b,\mu,\zeta\right)+i\epsilon_{\perp}^{\mu \nu}b_{\mu}S_{\perp \nu}M f_{1T,q/p}^{\perp}\left(x,b,\mu,\zeta\right)\,,  \label{eq:FullTMDb}
\end{align}
i.e., it includes contributions from both the unpolarized and the Sivers quark TMDs.
}

The dipole-target scatting matrix is given by:
\begin{equation}
    S_x\left({\mathbf{x},\mathbf{y}}\right) =   \frac{1}{N_{c}} \Tr \expval{ V\left(\mathbf{\mathbf{x}}\right) V^{\dagger}\left(\mathbf{y}\right)}_{x}\,,
\end{equation}
where $V(\mathbf{x})$ is the small-$x$ Wilson line  describing quark--target scattering. In writing Eq.~\eqref{eq:FullTMDb} we have assumed that one can factorize the impact-parameter dependence of the dipole--target scattering matrix and write
\begin{align}
    S_{x}\left(\mathbf{r}\right)B_{\perp} =\int \dd[2]{\mathbf{R}}   \frac{1}{N_{c}} \Tr \expval{ V\left(\mathbf{R+\frac{r}{2}}\right) V^{\dagger}\left(\mathbf{R-\frac{r}{2}}\right)}_{x} \,,\label{eq:IntegratedDipAmp}
\end{align}
where $\mathbf{R}$ is the impact parameter and $B_{\perp}$ is the average transverse area of the target hadron.
We can also decompose the $S$-matrix into real and imaginary parts,
\begin{align}
        S_{x}\qty(\mathbf{r}) =P_{x}\qty(\abs{\mathbf{r}})
        +i\epsilon_{\perp}^{\mu \nu} r_{\mu}S_{\perp \nu}MO^{\perp}_{1T,x}\qty(\abs{\mathbf{r}}) , \label{eq:FullDipoleAmp}
    \end{align}
in analogy to the quark TMD decomposition in Eq.~\eqref{eq:FullTMDb}. The real part $P_x$ of the amplitude is commonly referred to as the pomeron amplitude since it corresponds to a C-even gluon exchange.
Similarly, the imaginary part $O^{\perp}_{1T, x}$ is the spin-dependent odderon which corresponds to a C-odd gluon exchange~\cite{Hatta:2005as}. Substituting Eq.~\eqref{eq:FullDipoleAmp} into Eq.~\eqref{eq:TMD_x_expanded}, we obtain separate expressions for the unpolarized TMD and the Sivers function in the small-$x$ regime at the initial scale $\mu_0^2=\zeta_0 = \mu_{b_*}^2$: 
\begin{align}
    f_{q/p}(x,b,\mu_{b_{*}},\mu_{b_{*}}^{2}) \overset{\text{small } x}{=} \frac{N_{c}B_{\perp}}{4\pi^4}\frac{1}{x}\int \dd[2]{r} \dd{\epsilon_{f}^{2}}\mathcal{K}\left(\bold{b},\bold{r},\epsilon_f\right)
\left[ 1-P_{x}\left(\abs{\mathbf{r}}\right)
   \right] \, ,\label{eq:TMDunpol_x_expanded}
\\  f_{1T,q/p}^{\perp}\left(x,b,\mu_{b_{*}},\mu_{b_{*}}^{2}\right) \overset{\text{small } x}{=}  \frac{N_{c}B_{\perp}}{4\pi^4}\frac{1}{x}\int \dd[2]{r} \dd{\epsilon_{f}^{2}}\mathcal{K}\left(\bold{b},\bold{r},\epsilon_f\right)
    \frac{\mathbf{b} \vdot \mathbf{r} }{\mathbf{b}^2}\, O^{\perp}_{1T,x}\qty(\abs{\mathbf{r}})
    .\label{eq:Sivers_x_expanded}
\end{align}

Here we arrived at this compact expressions by noting that Eq.~\eqref{eq:Kernel} is even under shift $\bold{r}\rightarrow-\bold{r}-\bold{b}$ and that the real part of the dipole-target scattering matrix is even in the sign of the transverse separation of the quark-antiquark dipole, while the imaginary part is odd. This is a manifestation of the fact that the imaginary part of the dipole-target scattering matrix is C-odd~\cite{Hatta:2005as} since flipping the sign of the transverse separation of the dipole amounts to swapping a quark for an antiquark. 

Finally, the evolved unpolarized quark TMD and the Sivers function in the small-$x$ regime at the scale $\mu$ and $\zeta$ are given by
\begin{align}
    f_{q/p}(x,b,\mu,\zeta) &= f_{q/p}(x,b,\mu_{b_{*}},\mu_{b_{*}}^{2}) \exp[-\frac{g_{2}}{2}\ln{\frac{\sqrt{\zeta}}{Q_{0}}}\ln\frac{b}{b_{*}}]\exp[-S_{\pert}(\mu,\mu_{b_{*}},\zeta)]\, ,\label{eq:EvolvedTMDunpol}\\
    f_{1T,q/p}^{\perp}(x,b,\mu,\zeta) &= f_{1T,q/p}^{\perp}(x,b,\mu_{b_{*}},\mu_{b_{*}}^{2}) \exp[-\frac{g_{2}}{2}\ln{\frac{\sqrt{\zeta}}{Q_{0}}}\ln\frac{b}{b_{*}}]\exp[-S_{\pert}(\mu,\mu_{b_{*}},\zeta)]\,,
    \label{eq:EvolvedSivers}
\end{align}
where the input TMDs $f_{q/p}(x,b,\mu_{b_{*}},\mu_{b_{*}}^{2})$ and $f_{1T,q/p}^{\perp}(x,b,\mu_{b_{*}},\mu_{b_{*}}^{2})$ at the initial scale are given by Eqs.~\eqref{eq:TMDunpol_x_expanded} and~\eqref{eq:Sivers_x_expanded}. Note that in our modeling, we include only the non-perturbative part of the CS evolution kernel which governs the $\zeta$-evolution at large $b$. The intrinsic transverse momentum contributions from the proton are assumed to be already encoded in the small-$x$ dipole distribution~\cite{Tong:2022zwp}.

An important subtlety is that the quark Sivers function, when expanded at small $x$, is C-odd:
\begin{align}
f_{1T,\bar{q}/p}^{\perp}\left(x,b,\mu_{b_{*}},\mu_{b_{*}}^{2}\right) = -f_{1T,q/p}^{\perp}\left(x,b,\mu_{b_{*}},\mu_{b_{*}}^{2}\right)\,,
    \label{eq:quarkSiversSign}
\end{align}
implying a sign difference between quark and antiquark contributions. In contrast, for the unpolarized case, the quark and antiquark TMDs are equal in the small-$x$ expansion:
\begin{align}
f_{\bar{q}/p}\left(x,b,\mu_{b_{*}},\mu_{b_{*}}^{2}\right) = f_{q/p}\left(x,b,\mu_{b_{*}},\mu_{b_{*}}^{2}\right)\,.
\label{eq:quarkTMDnosign}
\end{align}
From the TMD perspective, this structure reflects the fact that the CGC dipole amplitude corresponds to a dipole-type gluon TMD. It has been shown that the T-odd dipole gluon distribution—i.e., the dipole gluon Sivers function—is C-odd. At the tree level, this can be understood by noting that its first transverse moment is related to the three-gluon correlation function in a transversely polarized proton, which are matrix elements of C-odd operators~\cite{Ji:1992eu,Beppu:2010qn,Dai:2014ala,Boer:2015vso,Benic:2024fbf,Boer:2015pni}. The C-odd nature of the small-$x$ expansion of the quark Sivers function has also been discussed in Refs.~\cite{Dong:2018wsp,Mantysaari:2025mht}. We can furthermore verify these relations explicitly by using:
\begin{equation}
     f^{[\gamma^+]}_{\bar q,\lambda}(x,\kt)
     =
     -f^{[\gamma^+]}_{ q,\lambda}(-x,-\kt),
\end{equation}
in Eq.~\ref{eq:TMDunpol_x_expanded}  and Eq.~\ref{eq:Sivers_x_expanded}, which is the (momentum space) relation between the quark and antiquark TMD PDFs.

To perform phenomenological studies in the small-$x$ regime using the above formulas, we require a model for the dipole $S$-matrix. We achieve this by specifying an initial condition at $x = 0.01$ and evolving it to smaller $x$ values using the Balitsky--Kovchegov (BK) equation~\cite{Balitsky:1995ub,Kovchegov:1999yj}:
\begin{align}
    \frac{\partial}{\partial \log(1/x)}S_{x}\left(\mathbf{r}\right) = \int \dd[2]{\mathbf{r'}}\mathcal{K}\left(\mathbf{r,r'}\right)\left[ S_{x}\left(\mathbf{r'}\right)S_{x}\left(\mathbf{r-r'}\right)-S_{x}\left(\mathbf{r}\right) \right]
\end{align}
where we use the Balitsky prescription~\cite{Balitsky:2006wa} for the BK kernel $\mathcal{K}$:
\begin{multline}
    \mathcal{K}\left(\mathbf{r,r'}\right) = \frac{N_{c}\alpha_{s}\left(r^2\right)}{2\pi^2}
    \left[\frac{\mathbf{r}^2}{\mathbf{r}^{\prime2}\left(\mathbf{r}-\mathbf{r}'\right)^{2}} +\frac{1}{\mathbf{r}^{\prime2}}\left(\frac{\alpha_{s}\left(\mathbf{r}^{\prime2}\right)}{\alpha_{s}\left(\left(\mathbf{r}-\mathbf{r}'\right)^2\right)} -1\right) \right. \\
    \left.+\frac{1}{{\left(\mathbf{r}-\mathbf{r}'\right)}^{2}}\left(\frac{\alpha_{s}\left(\mathbf{r}^{\prime 2}
    \right)}{\alpha_{s}\left(\left(\mathbf{r}-\mathbf{r}'\right)^2\right)}-1 \right)\right]\,.
\end{multline}
The coordinate-space running coupling $\alpha_s(r^2)$ is defined as:
\begin{align}
    \alpha_{s}\left(r^{2}\right)  = \frac{12\pi}{\left(33-2N_{f}\right)\log{\left(\frac{4C^{2}}{r^{2}\LambdaQCD^2}
    \right)}}\,
\end{align}
with $N_f = 3$, $\LambdaQCD = \SI{0.241}{GeV}$, and $C^2$ is a parameter describing the connection between coordinate and momentum space expressions of the running coupling.

For the initial condition of the BK evolution, we use the MVe model~\cite{Lappi:2013zma} for the real part (pomeron component) of the $S$-matrix:
\begin{align}
    P_{x}(r) = \exp\left[-\frac{r^{2}Q^{2}_{s0}}{4}
    \log\left(\frac{1}{r\LambdaQCD}+e_{c}\cdot e\right) \right] \,.
    \label{eq:MvModel}
\end{align}
The imaginary part (odderon component) is generally less constrained. Following Refs.~\cite{Lappi:2016gqe,Yao:2018vcg}, we relate it to the real part via
\begin{equation}
      O^{\perp}_{1T,x}(r) = - P_{x}\qty(r)\kappa \frac{r^2Q_{s0}^{3}}{8 M_p} \label{eq:Odderon}
\end{equation}
where $\kappa = 1/3$. For the free parameters, we use the values from the fit in Ref.~\cite{Casuga:2023dcf}:
\begin{align}
    C^2 &= 4.97,
    &
    e_c &= 35.3,
    &
    Q^2_{s0} &= \SI{0.061}{GeV^2},
    &
    B_\perp &= \SI{14.1}{mb}.
\end{align}

\begin{figure}[t]
  \centering
  \subfigure[Unpolarized quark TMD]{%
    \includegraphics[width=0.47\textwidth]{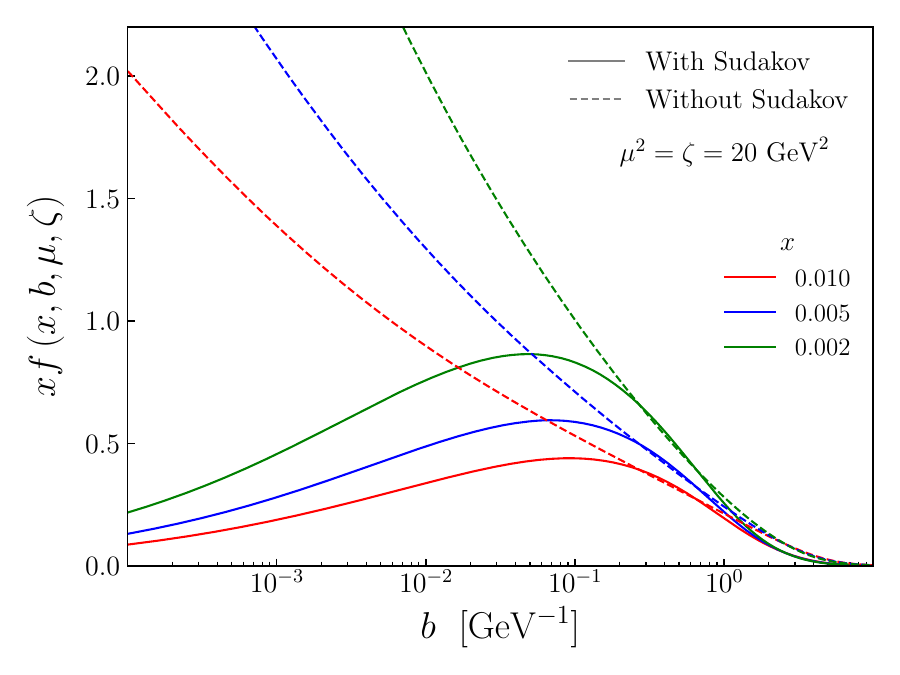}
   \label{fig:unpolarized_TMD}
  }
  \subfigure[Sivers quark TMD]{%
    \includegraphics[width=0.47\textwidth]{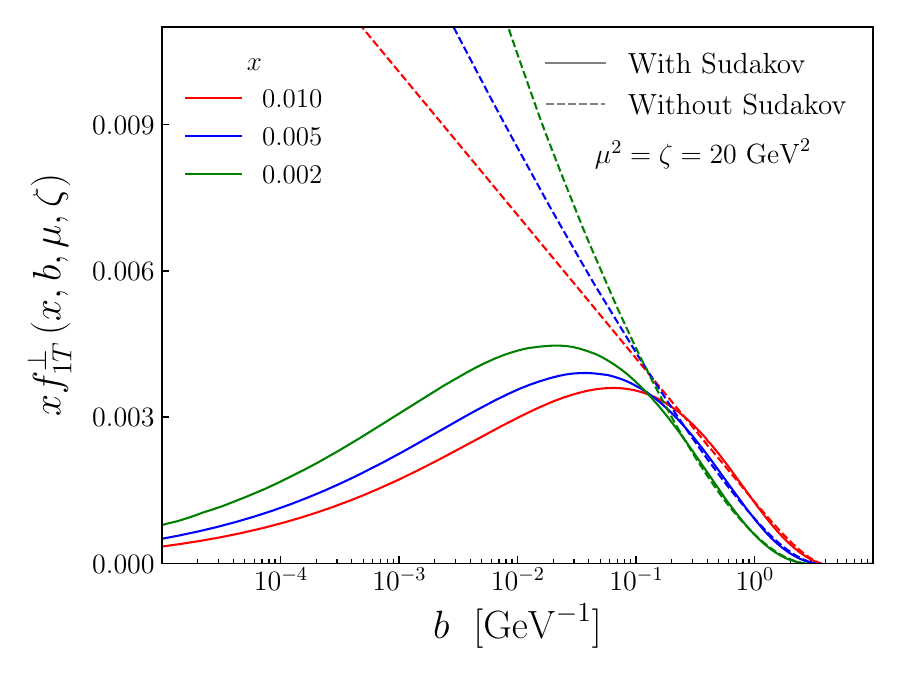}
    \label{fig:polarized_TMD}
  }
  \caption{
  The quark TMDs from matching to CGC, with and without the Sudakov terms in Eqs.~\eqref{eq:EvolvedTMDunpol} and \eqref{eq:EvolvedSivers}.
  }
  \label{fig:two_plots}
\end{figure}

To illustrate the behavior of our model for the quark TMDs, we plot them as a function of $b$ for different values of $x$ in Fig.~\ref{fig:two_plots}.
Evolving the distributions to smaller $x$,
we see that the unpolarized quark TMD generally increases,
whereas the quark Sivers TMD  increases for $b \lesssim \SI{0.1}{GeV^{-1}}$
and decreases for $b \gtrsim \SI{0.1}{GeV^{-1}}$.
This trend is in line with the small-$x$ dependence of the corresponding pomeron and odderon contributions~\cite{Lappi:2016gqe}.  In particular, the turnover dependence of the energy dependence of the quark-Sivers function is analogous to the turnover of the energy dependence of the odderon amplitude in \cite{Yao:2018vcg}. For comparison, we also show the quark TMDs without the Sudakov factors---the non-perturbative $g_2$ term and the perturbative component $S_{\text{pert}}$---appearing in Eqs.~\eqref{eq:EvolvedTMDunpol} and~\eqref{eq:EvolvedSivers}. Since these Sudakov terms arise from the TMD evolution, our results show that the TMD evolution suppresses both quark TMDs for small values of $b$, changing the limiting behavior $b \to 0$ quite drastically.
Without the TMD evolution, the  $b\rightarrow0$ asymptotic behavior of the unpolarized quark TMD in our model comes from the region with dipole sizes $b\ll r  \ll Q_s^{-1}$ which yields $xf \sim  \int^{}_b\frac{dr}{r}\,\ln{\frac{1}{r\Lambda_{\mathrm{QCD}}}} \sim \left(\log{b}\right)^2$, consistent with the numerical result in Fig.~\ref{fig:unpolarized_TMD}. In the same limit the leading asymptotic behavior for the quark Sivers TMD is instead $xf_{1T}^{\perp}\sim \int_b \frac{dr}{r}\sim -\log{b}$, which is again consistent with the numerical result in Fig.~\ref{fig:polarized_TMD}.
Including the TMD evolution largely suppresses both quark distributions at small $b$.

\subsection{TEEC Jet Function \label{sec:Jetfun}}

The TMD fragmentation functions have been extracted from global fits to semi-inclusive DIS and Drell--Yan data; see, e.g., Refs.~\cite{Bacchetta:2024qre,Moos:2025sal}. In this work, we follow the model used in Refs.~\cite{Echevarria:2020hpy,Sun:2014dqm,Kang:2024otf,Kang:2023oqj} and write the TEEC jet function as
\begin{multline}
    J_{\mathbb{S}/q}\left(b,\mu,\zeta'\right)  =\sum_{h\in \mathbb{S}}\int_{0}^{1}\dd{z}zD_{1,h/q}\left(z,\mu_{b_{*}}\right) \\ \times
 \exp\left[-S^{D}_{\text{NP}}\left(z,b,Q_{0},\zeta'\right)\right]\exp\left[-S_{\text{pert}}\left(\mu,\mu_{b^{*}},\zeta'\right)\right] \,, \label{eq:Evolved jetfunction}
\end{multline}
where we retain only the leading-order matching coefficient in the perturbative expansion of the TMD fragmentation function. Here, the perturbative Sudakov factor $S_\pert$ is the same as that for the quark TMDs and is given in Eq.~\eqref{eq:perturbative_Sudakov}. The non-perturbative Sudakov factor $S_\np^{D}$ is modeled as
\begin{align}
S^{D}_{\text{NP}}\left(z,b,Q_{0},\zeta'\right)=\frac{g_{2}}{2}\ln{\left(\frac{b}{b_{_{*}}}\right)}\ln{\qty(\frac{\sqrt{\zeta'}}{Q_{0}})}+g_{1}^{D}\frac{b^{2}}{z^{2}}\,,
\end{align}
with $g_{1}^{D}=\SI{0.042}{GeV^{2}}$~\cite{Sun:2014dqm}.

Note that if we were to choose $\mathbb{S}=\text{\{all hadrons\}}$, the jet function for quarks and antiquarks would be equal: $J_{\bar q} = J_q$. In this case, due to the opposite signs of the quark and antiquark Sivers functions in Eq.~\eqref{eq:quarkSiversSign}, the Sivers contribution to the TEEC, $\Sigma^{UT}_{\mathbb{S}}$ vanishes identically in the small-$x$ limit, as can be seen directly from the factorized expression in Eq.~\eqref{eq:TEEC_polarized}.

For this reason, it is necessary to restrict the TEEC measurement to a subset of the final-state hadrons such that the quark and antiquark jet functions are no longer equal. One such case that we consider in this work is restricting the final-state hadrons to either positively or negatively charged hadrons. The importance of measuring events with charged hadrons in the final state to probe small-$x$ Sivers asymmetry has also been noted in Refs.~\cite{Mantysaari:2025mht,Boer:2015pni}.
Alternatively, one can introduce charge tracks to the TEEC definition such that each hadron is weighted by its electric charge~\cite{Lee:2023npz,Lee:2023tkr}.
This effectively cancels the relative sign between the quark and antiquark Sivers functions, allowing their contributions to add coherently and resulting in a non-zero Sivers asymmetry in the small-$x$ regime.

Finally, note that charge conjugation relates the jet function of a quark fragmenting into a positively charged hadron to that of the corresponding antiquark fragmenting into a negatively charged hadron:
\begin{equation}
    J_{h^{+}/q}\left(b,\mu,\zeta'\right) = J_{h^{-}/\bar{q}}\left(b,\mu,\zeta'\right)\,,
\end{equation}
and similarly $J_{h^{-}/q} = J_{ h^{+} /\bar q}$.
This symmetry also implies that the Sivers asymmetries are equal in magnitude but opposite in sign:
\begin{equation}
    A^{h^{+}}_{UT} = -A^{h^{-}}_{UT}\,,
    \label{eq:AUT-charge}
\end{equation}
i.e., the sign of the Sivers asymmetry depends on whether one focuses on positively or negatively charged hadrons. This relation can be readily verified from the factorized expressions for $\Sigma^{UU}_{\mathbb{S}}$ and $\Sigma^{UT}_{\mathbb{S}}$ given in Eqs.~\eqref{eq:TEECsub} and \eqref{eq:TEEC_polarized}, respectively---together with the symmetry properties of the quark and antiquark contributions: $f_{\bar q/p} = f_{q/p}$ and $f_{1T,\bar q/p}^\perp = -f_{1T,q/p}^\perp$---as shown previously in Eqs.~\eqref{eq:quarkTMDnosign} and \eqref{eq:quarkSiversSign}, along with the charge-conjugation relations for the jet functions discussed above.

\begin{figure}[t]
\centering
\includegraphics[width=0.7\textwidth]{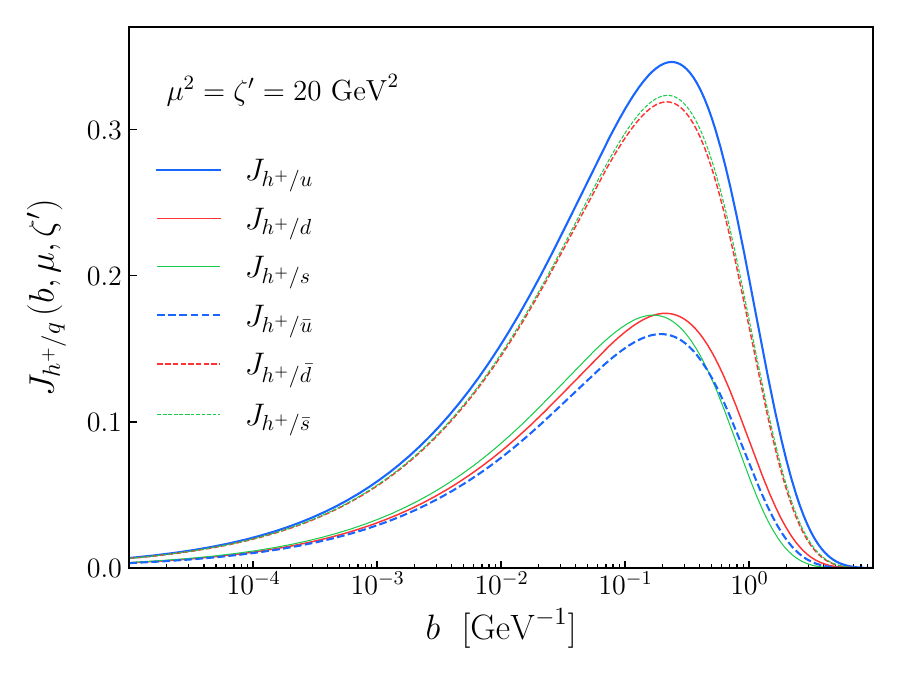}
   \caption{\label{fig:jfuns}TEEC jet functions, $J_{h^{+}/q}$, (Eq.~\eqref{eq:Evolved jetfunction}) for positively charged hadrons plotted at the scale $\mu^2 = \zeta'  = \SI{20}{GeV^2}$.}
\end{figure}

For the numerical implementation, we need a model for the jet function in Eq.~\eqref{eq:Evolved jetfunction}.
We choose the following simple parametrization:
\begin{equation}
    \sum_{h\in\mathbb{S}}\int_{0}^{1}\dd{z} zD_{1,h/q}\left(z,\mu_{b_*}\right)\text{exp}\left(-g_{1}^{D}\frac{b^2}{z^2}\right) = N_{q}\exp\left(-g_{q}b\right) \,,
    \label{eq:fit}
\end{equation}
where $N_q$ and $g_q$ are free parameters.
The normalization constant $N_q$ is necessary because
\begin{equation}  \sum_{h\in\mathbb{S}}\int_{0}^{1}\dd{z}zD_{1,h/q}\left(z,\mu_{b}\right)<1,
\end{equation}
unless $\mathbb{S} = \text{\{all hadrons\}}$ in which case the momentum-sum rule guarantees $N_q = 1$ identically. Using the parametrization in Eq.~\eqref{eq:fit}, we perform a fit with the NPC23 collinear fragmentation functions for charged hadrons~\cite{Gao:2024dbv}.
In particular, the fit is carried out using FFs for positively charged hadrons. The coefficients obtained are listed in Table~\ref{tb:coeff}. The resulting TEEC jet functions are shown in Fig.~\ref{fig:jfuns}. As expected,  the jet function $J_{h^{+}/q}$ is larger when the fragmenting parton carries a positive charge ($u$, $\bar d$, or $\bar s$). This can be understood from the valence quark structure of positively charged hadrons.

\begin{table}[t]
\centering
\begin{tabular}{|c|c|c|}
\hline
Fragmenting quark & $N_{q}$ & $g_{q} \ \text{[GeV]}$ \\
\hline
$u$ & 0.453 &0.810 \\
$\bar{u}$ &0.224 & 1.122 \\
$d$ & 0.237 & 0.981 \\
$\bar{d}$ & 0.430 & 0.942 \\
$s$ & 0.254 &1.387 \\
$\bar{s}$ & 0.434 & 0.921 \\
\hline
\end{tabular}
\caption{Resulting parameters from fitting the functional form of Eq.~\eqref{eq:fit} to NPC23 fragmentation functions of light quarks into positively charged hadrons.
}
\label{tb:coeff}
\end{table}

\section{Sivers asymmetry} \label{sec:Pheno}
With the models for the quark TMDs and the TEEC jet function outlined in detail in the previous section, we are now ready to compute the TEEC Sivers asymmetry in the small-$x$ regime, as defined in Eq.~\eqref{eq:Asymmetry}, for the case $\mathbb{S} = \{\text{positively charged hadrons}\}$, i.e., $A^{h^{+}}_{UT}$. This observable provides a direct probe of the spin-dependent odderon at small $x$. As discussed earlier, the corresponding asymmetry for $\mathbb{S} = \{\text{negatively charged hadrons}\}$ is equal in magnitude but opposite in sign, as given in Eq.~\eqref{eq:AUT-charge}.

\begin{figure}[t]
\centering
\includegraphics[width=0.7\textwidth]{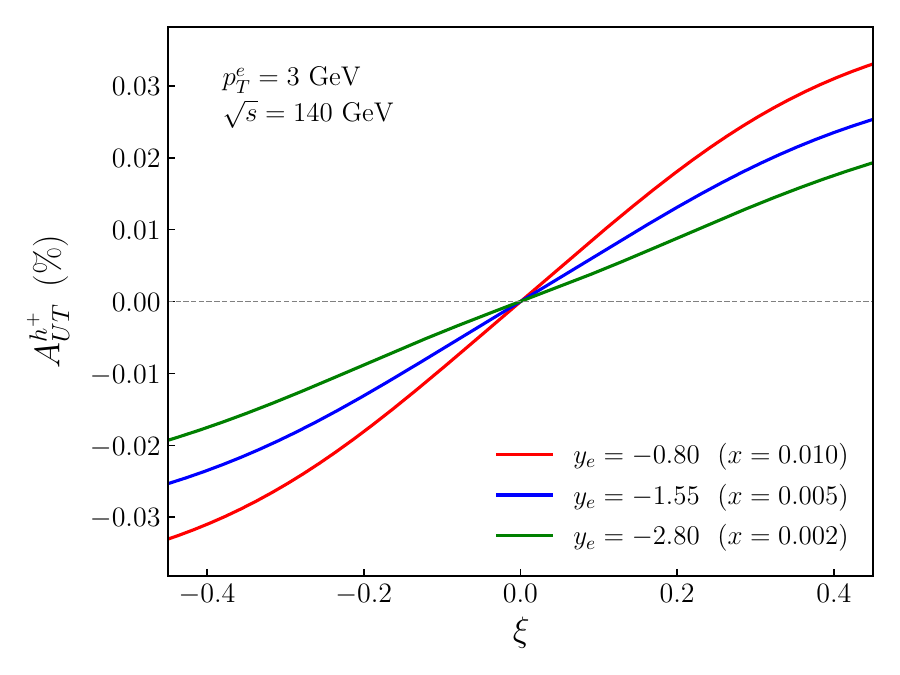}
   \caption{\label{fig:Asym}
   Predicted TEEC Sivers asymmetry at the EIC for a transversely polarized proton with positively charged hadrons as a function of the \update{$\xi$} variable.
   Different colors correspond to different values of the outgoing electron rapidity $y_e$ or, equivalently, different values of the Bjorken $x$ variable.
   }
\end{figure}

We work at a fixed center-of-mass energy $\sqrt{s} = \SI{140}{GeV}$, corresponding to the highest projected energy at the future Electron-Ion Collider, and choose kinematics that probe different values of $x$ such that the small-$x$ description remains applicable. The resulting predictions are presented in Fig.~\ref{fig:Asym}, where the TEEC Sivers asymmetry $A^{h^{+}}_{UT}$ is shown as a function of $\xi$. We find that the asymmetry is largest at the initial condition of the BK evolution ($x = 0.01$) and becomes progressively suppressed as $x$ decreases. This behavior can be attributed to the BK evolution, which reduces the odderon contribution at smaller $x$~\cite{Lappi:2016gqe,Yao:2018vcg}, and is consistent with the trend observed for the quark Sivers function in the small-$x$ regime, as shown in Fig.~\ref{fig:two_plots}.

We also observe that the asymmetry increases with increasing $|\xi|$, i.e. when we are going away from the back-to-back limit.
This is due the polarized TEEC in Eq.~\eqref{eq:TEEC_polarized} having $\sin{\left(\xi bp_{T}^{e}\right)}$ as opposed to $\cos{\left(\xi bp_{T}^{e}\right)}$ in the unpolarized case in Eq.~\eqref{eq:TEECsub}, leading to an asymptotic small-$\xi$ behavior $A^{h^+}_{UT} \sim \xi$.
This increase of the asymmetry seems to slow down eventually when going away from the back-to-back limit, at $|\xi| \gtrsim 0.4$, although we cannot fully assess this region in our formalism based on the TMD factorization that assumes $|\xi| \ll 1$. In general, we find the asymmetry to be of the order $\order{0.01 \%}$ in the back-to-back limit where our formalism is valid.

\section{Conclusions} \label{sec:Conclusion}

In this work, we have studied the Sivers effect in the context of the transverse energy--energy correlator (TEEC) for an approximately back-to-back hadron--electron pair in the center-of-mass frame of electron--proton collisions, where the proton is transversely polarized. Within the TMD framework, we express the TEEC observables for both unpolarized and polarized scattering in terms of quark TMDs and TEEC jet functions.

Furthermore, we consider the process in the small-$x$ kinematic regime, where the color glass condensate description becomes applicable, and expand the quark TMDs in terms of the small-$x$ dipole amplitude. This approach allows us to directly relate the Sivers effect to an odderon interaction with the target, which remains one of the least understood aspects of small-$x$ QCD dynamics. Due to the C-odd nature of the odderon, the quark and antiquark Sivers functions acquire opposite signs, resulting in a complete cancellation of the polarized TEEC in the fully inclusive case when summing over all final-state hadrons. To circumvent this, we restrict the final state to include only positively charged hadrons, thereby yielding a non-vanishing Sivers asymmetry at small $x$. Alternatively, one could achieve the same sensitivity by incorporating charge-weighted tracks into the TEEC definition~\cite{Lee:2023npz}, as discussed in, e.g., Ref.~\cite{Mantysaari:2025mht}.

As expected from its connection to the odderon interaction, we find that the Sivers asymmetry becomes progressively suppressed with decreasing Bjorken $x$. For $x \lesssim 0.01$, where the small-$x$ formalism remains applicable, the asymmetry is found to be at the level of $0.01\%$. This is two orders of magnitude smaller than results from other studies performed at larger $x$, such as in SIDIS~\cite{Echevarria:2020hpy} and in jet+$\text{J}/\psi$ production~\cite{Kishore:2019fzb} at EIC kinematics, as well as in Drell--Yan processes at SpinQuest (Fermilab)~\cite{Fernando:2023obn} and RHIC~\cite{Kang:2012vm}, all of which report asymmetries of order $\mathcal{O}(1\%)$. Our results are also approximately one order of magnitude larger than previous predictions for dijet production~\cite{Liu:2020jjv,Kang:2020xez} at RHIC kinematics.

We conclude that the TEEC in polarized electron--proton scattering provides a novel and complementary channel to probe the elusive odderon interaction in QCD.
As the overall size of the odderon contribution is largely unknown, even a crude measurement of the Sivers asymmetry would be enough to give us significant constraints on the odderon interaction.
Other proposals for accessing the odderon at the EIC have also been suggested~\cite{Mantysaari:2025mht,Benic:2024fbf,Benic:2024pqe}, and we expect that exploring a broad set of observables will be essential for advancing our understanding of this unique gluonic correlation.

\section*{Acknowledgment}
We thank Adrian Dumitru, Heikki Mäntysaari, and Marco Radici for discussions and comments. Z.K., D.P., and J.P. are supported National Science Foundation under grant No.~PHY-2515057. This work is also supported by the U.S. Department of Energy, Office of Science, Office of Nuclear Physics, within the framework of the Saturated Glue (SURGE) Topical Theory Collaboration.

\bibliographystyle{JHEP-2modlong.bst}
\bibliography{references}

\end{document}

%% file: packages.tex
\usepackage[utf8]{inputenc}

\usepackage{xcolor}

\usepackage{amsmath}
\usepackage{amssymb}
\usepackage{tikz}
\usepackage{siunitx}
\usepackage{physics}
\usepackage{slashed}

\usepackage{overpic}

\usepackage{graphicx}

\usepackage{mathtools}

%% file: definitions.tex
\newcommand{\kt}{{\mathbf{k}}}

\newcommand{\pert}{\text{pert}}
\newcommand{\np}{\text{NP}}
\newcommand{\LambdaQCD}{\Lambda_\text{QCD}}